\documentstyle[preprint,aps]{revtex}
\draft
\begin{document}
\title{Pion Transparency in 500 MeV $C(\pi,\pi^{'})$ Reactions?}
\bigskip
\author{\bf Bao-An Li$^a$, Wolfgang Bauer$^{b}$ and Che Ming Ko$^a$}
\address{a. Cyclotron Institute and Department of Physics,\\ 
Texas A\&M University, College Station, TX 77843, USA\\
b. Department of Physics and Astronomy,\\
and National Superconducting Cyclotron Laboratory,\\ 
Michigan State University, East Lansing, MI 48824-1321, USA}
\maketitle

\begin{quote}
The question whether there is a pion transparency in 500 MeV $C(\pi, \pi^{'})$
scatterings is studied using a semiclassical, hadronic transport model. 
The double differential cross sections of this reaction measured at LAMPF 
can be largely accounted for, if one uses energy-dependent, anisotropic 
angular distributions which are fitted to pion-nucleon scattering data
for the decay of $\Delta(1232)$ and $N^*(1440)$ resonances.
The remaining discrepancy between the data and the calculation sets a limit on 
effects of more exotic processes.
\end{quote}

\newpage
Pion$-$nucleus scatterings at energies near the peak of $\Delta(1232)$ 
formation ($T_{\pi}\approx 180$ MeV) have been studied extensively at
several meson facilities during the last two decades. Experiments at 
LAMPF, KEK and BNL using high energy pions far above the resonance peak 
have recently started to accumulate interesting data\cite{bnl,lanl1,lanl2,kek}.
Since the pion$-$nucleon cross section decreases by almost an order of 
magnitude as the pion energy increases from the resonance energy to 
higher energies, high energy pions are therefore 
expected to penetrate more deeply into the interior of nuclei. 
To understand the dynamics of these high energy pions in nuclear matter 
is one of the major goals of current experimental studies on pion$-$nucleus 
scatterings. These studies will be further enhanced by the 
availability of pion beams at GSI/Darmstadt, Germany in the near future.

Among these studies, one interesting phenomenon was recently found at LAMPF in
inclusive scatterings of 500 MeV pions from a carbon target\cite{lanl1}.
The ($\pi,\pi^{'}$) inclusive cross sections, with outgoing pion energies
near 200 MeV, failed to show the strong depletion predicted by an intranuclear
cascade model (INC)\cite{gibbs}. To bring the INC prediction close to the 
data requires that pions from the ($\pi,2\pi$) channel do not interact 
with the nucleus for about 2 fm/c. As an alternative mechanism for filling 
up this dip, the possibility of forming a $\sigma$ meson was also suggested.
Since the isospin zero character of the $\sigma$ prevents it from interacting 
strongly with the nucleus, it will mainly decay into two pions outside 
the nucleus. Both mechanisms thus require that the nuclear medium is 
transparent to pions that are from the $(\pi,2\pi)$ production channel. 
It is also interesting to mention that the Relativistic Quantum Molecular 
Dynamics (RQMD 1.08) has been used as well to understand the data\cite{john}. 
However, it underestimates the data at forward angles by a factor of 2-7 in 
the whole energy range. At large angles, it overestimates the data by a 
factor of 2-6 for energy losses between 200 and 300 MeV and underestimates
significantly other parts of the spectra. The discrepancies between the 
data and the model predictions, as well as the different mechanisms 
proposed to explain them, have attracted much attention 
and stimulated many interesting discussions. This 
problem was assigned to several participants as a homework by the organizers of 
the workshop on ``pionic processes and transport in hadronic matter'' 
at Los Alamos in July, 1995\cite{las}. In this Letter we report results 
of our study using a hadronic transport model BUU/ART\cite{buu,art}. 

The BUU/ART model has been used successfully in studying many aspects 
of heavy-ion collisions from low to relativistic energies, especially the pion
observables (e.g. \cite{liko}). Since pions are the most copiously 
produced particles in relativistic heavy-ion collisions, it is crucial 
to understand very well their transport dynamics in nuclear matter. 
In this respect pion$-$nucleus collisions provide a clean testing ground. 
It is therefore interesting to test models for heavy-ion collisions 
against data from pion$-$nucleus collisions. In fact, a wealth of
interesting physics 
on the propagation of pions has been extracted from pion$-$nucleus 
scatterings during the last two decades. In the present letter, we have 
incorporated most of these effects into our semiclassical transport 
model in order to compare with the recent data from pion induced 
reactions.
  
Three aspects of the BUU/ART model, pion-scattering, -production 
and -absorption, are important for studying pion$-$nucleus interactions. 
In this model, pion scatterings are modeled through the formation and decay of 
$\Delta(1232)$ and $N^{*}(1440)$ resonances. Although these resonances 
should decay in definite angular momentum channels ($P_{33}$ and $P_{11}$),
in heavy-ion collisions it is customary to assume that baryon resonances 
decay isotropically in their center of mass frames since frequent 
rescatterings of these resonances in dense hadronic matter largely smear 
out information about the angular distribution from their decays. 
For describing pion$-$nucleus scatterings, however, it is important to 
incorporate properly experimentally measured angular distributions 
in pion$-$ nucleon scatterings. These distributions were found to be
strongly energy dependent\cite{data}. More specifically, the distribution 
is about forward-backward symmetric only at center of mass energies near 
the peak of the delta resonance (1232 MeV) due to P-wave dominance. 
It is strongly backward peaked at lower energies, and forward 
peaked at higher energies due to the contribution and interference of 
many partial waves. Within the present model our classical treatment has 
no way of reproducing naturally the interference and thus the experimental 
angular distribution. To evaluate effects of the pion$-$nucleon 
angular distribution on pion double differential cross sections in 
pion$-$ nucleus scatterings, we will compare two extreme 
cases by forcing the resonances to decay either isotropically or 
according to the measured pion$-$nucleon angular distributions. 
It is worth mentioning that our treatment here is rather
similar in spirit to those of the early INC models for pion induced 
reactions\cite{harp,gin}.

The measured pion$-$nucleon angular distribution in their center of mass 
frame can be expressed in terms of Legendre polynomials 
frame\cite{marshak,gia}
\begin{equation}\label{dif}
\frac{d\sigma}{d\Omega^*}=\sum_{n=0}^{n_{max}}c_nP_n(cos\theta^*).
\end{equation}
The coefficients $c_n$ at different energies are listed in ref.\ \cite{data} 
for $\pi^{\pm}p$ scatterings. We use proper Clebsch-Gordon coefficients to 
obtain the angular distributions for the decay of resonances with different 
charges. Shown in Fig.\ 1 are the angular distributions for the decay of 
$\Delta^{++}$ resonances having masses of 1.1, 1.23, 1.5 and 1.8 GeV.
 
At $T_{\pi}=500$ MeV the total $\pi+N$ inelastic
cross section results almost completely from the production of two 
pions \cite{data1}, $\pi+N\rightarrow 2\pi+N$. 
We model this process by using the reactions $\pi+N\rightarrow 
\pi+\Delta$ and $\pi+N\rightarrow \rho+N$. Assuming the formation of higher 
resonances ($\Delta^{*}$ and $N^{*}$) in the intermediate state\cite{art}, 
the branching ratio of the two channels can be estimated according to
\begin{equation}
\frac{\sigma(\pi+N\rightarrow \pi+\Delta)}{\sigma(\pi+N\rightarrow \rho+N)}
=\frac{\sum_i(2J_i+1)\Gamma_i(R_i\rightarrow \pi\Delta)W(R_i)}
{\sum_i(2J_i+1)\Gamma_i(R_i\rightarrow \rho N)W(R_i)},
\end{equation}
which is about a constant of 1.7 for $1.22\leq \sqrt{s}_{\pi N}\leq 1.66$ GeV. 
In the above, $R_i$ are the higher $\Delta^*$ and $N^*$ resonances with 
masses up to 1.7 GeV, $\Gamma_i(R_i\rightarrow \pi \Delta (\rho N)$ are the 
partial decay widths of these resonances and 
\begin{equation}
W(R_i)=\frac{m_i^2\Gamma(\pi N\!\rightarrow\!R_i)}
            {(m_i^2-s)^2+m_i^2\Gamma^2(\pi N\!\rightarrow\!R_i)}
\end{equation}
is  the Breit-Wigner formula for the formation of these resonances. 

Pion absorption is modeled through either
the two-step, two-body processes $\pi+N\rightarrow \Delta(N^*)$ and 
$\Delta(N^{*})+N\rightarrow N+N$ or the quasi-deuteron absorption 
processes $\pi+D\rightarrow NN$. Parameters used for the two-body, two-step 
processes have been discussed in detail in ref.\ \cite{buu}. For the 
quasi-deuteron process $\pi^++np\rightarrow pp$ we use the experimental 
cross section \cite{ritchie} 
\begin{equation}\label{qd}
\sigma(\pi^++np\rightarrow pp)=\frac{b}{\sqrt{T_{\pi}}}
+\frac{c}{(\sqrt{s}_{\pi np}-E_R)^2+d}+a;
\end{equation}
with $a=-1.2$ mb, $b=3.5$ mb\,MeV$^{-1/2}$, $c=7.4\cdot 10^4$ mb\,MeV$^{-2}$,
$d=5600$ MeV$^2$ and
$E_R=2136$ MeV. In the above $\sqrt{s}_{\pi np}$ is the center of mass energy
of the $\pi-D$ system. The 
quasi-deuteron is identified as two nearest nucleons with a maximum separation 
of 3 fm and proper charges to match the pion. Cross sections for pion 
absorptions on other quasi-deuterons are obtained via\cite{engel} 
\begin{eqnarray}
\sigma(\pi^-np\rightarrow nn)&=&\sigma(\pi^+np\rightarrow pp),\\            
\sigma(\pi^0np\rightarrow np)&=&0.44\cdot\sigma(\pi^+np\rightarrow pp),\\
\sigma(\pi^0pp\rightarrow pp)&=&\sigma(\pi^0nn\rightarrow nn)=0.14\cdot
\sigma(\pi^+np\rightarrow pp),\\
\sigma(\pi^+nn\rightarrow np)&=&\sigma(\pi^-pp\rightarrow np)=
0.083\cdot\sigma(\pi^+np\rightarrow pp).
\end{eqnarray}
The three terms in Eq. (\ref{qd}) represent respectively contributions 
to the quasideutron absorption from s-wave, p-wave and higher partial 
waves as well as the interference among them. The quasideutron absorption 
process allows us to include the absorption of low energy pions via s-wave 
properly. It may also introduce some double counting of the
p-wave absorption. We estimate, however, that this double counting is very 
small as the second term in Eq.\ (\ref{qd}) has a maximum of only about 10 mb 
which is very small compared to that in the two-body, two-step process.   
On the other hand, it also seems improper to simply throw away the p-wave
contribution in Eq.\ (\ref{qd}) as our model does not include the interference
of different partial waves. To minimize possible double countings of the 
p-wave absorption we allow a pion and nucleons to participate in only 
one of the pion absorption channels during each time step. Moreover, 
the quasi-deutron absorption process is treated explicitly as a one-step 
process.

In transport model calculations, we need to know the initial distribution 
of nucleons in the nucleus. For a carbon nucleus, 
nucleons are initialized using the experimentally
measured charge distribution
\begin{equation}
   \rho(r)/\rho_0 = (1+wr^2/c^2) \left(1+\exp((r-c)/z)\right)
\end{equation}
with $c=2.36$ fm, $z=0.522$ fm and $w=-0.149$.
The momenta of nucleons are assigned using the local Thomas-Fermi 
approximation.

To test the validity of the transport model, 
we first perform a study on pion absorption cross sections in pion$-$nucleus
scatterings in a large beam energy range. This study is a prerequiste
for us to study the double differential cross sections in the following. 
We notice that the pion absorption cross sections in pion$-$nucleus 
scatterings have been studied in a similar transport model in 
ref.\ \cite{engel} for various targets. Here we perform a study 
for $\pi^++^{12}C$ reactions at 
beam energies from 50 to 500 MeV in Fig.\ 2. The data are taken from ref. 
\cite{ashery} (filled circles), ref. \cite{jones} (filled squares) and 
ref. \cite{ransome} (fancy squares). Results using the ``standard'' 
BUU/ART model (i.e.\ without changes in the elementary cross sections 
described above, and assuming isotropic decays of resonances in 
their rest frames) are shown with the open circles. The line is drown 
to guide the eye. It is seen that the calculations agree reasonably 
well with the data within error bars. 

In Fig.\ 3 the inclusive double differential spectra 
from the scattering of 500 MeV $\pi^{+}$ on carbon are shown as a 
function of the energy loss ($T_{in}-T_{out})$. The symbols with error 
bars are the data and the lines are from calculations. 
The results from the ``standard'' BUU/ART model calculations are given by 
the dotted lines. First, in contrast to the INC predictions,
we do not see the one-order-of-magnitude depletion in cross sections 
from the data at energy losses around 
300 MeV at forward angles. Second, we notice that the calculations 
underestimate significantly the quasielastic peaks at 30, 40 and 50 degrees, 
and also overestimate the data at 70, 90 and 110 degrees. Third, the elastic 
peaks at forward angles corresponding to the recoil of the whole carbon 
nucleus can not be reproduced by the model as one expects.

We find that the agreement with the data can be much improved by allowing 
the resonances to decay anisotropically according to Eq.\ (\ref{dif}). 
This result is shown by the solid lines. It is fair to say that the quality 
of fit to the data is at about the same level as the INC calculations 
\cite{gibbs} using a 2 fm/c formation time for pions from the $(\pi, 2\pi)$ 
process. From this one might conclude that the evidence for the formation 
time effects previously reported is not yet conclusive.

It is interesting to see that the improvement due to the anisotropical 
angular distributions is mainly around the quasielastic peaks. This is 
because the high energy-loss parts of the spectra are due to pions 
produced in $(\pi, 2\pi)$ process, which is in agreement with the 
finding of the INC calculations. We notice that in both the INC and the 
present model the calculated spectra at 30, 40 and 50 degrees are smaller 
than the data by about a factor of two in the energy range of 
$T_{in}-T_{out}\geq $ 150 MeV. This discrepancy may be used to 
limit effects of possible exotic processes. We have not yet tested the 
possibility of forming the $\sigma$ meson in the model. On the other hand, 
we have studied effects of possible enhancement of the in-medium 
pion$-$nucleon inelastic cross sections. This is motivated by studies 
in $K^+$-nucleus scatterings where the underestimate of 
theoretical calculations compared to the experimental data 
has led to the suggestion that the kaon-nucleon cross section is 
enhanced in the nuclear medium \cite{paez,siegel,chen,brown}. Possible 
explanations for the increased cross section are the increase in the nucleon's 
size in medium due to QCD related physics \cite{siegel}, or the scattering 
from enhanced meson clouds caused by the reduction of meson 
masses in medium\cite{brown}. Since high energy pions have mean 
free paths compatible to kaons, signatures of these
exotic processes were expected to be identifiable in high energy pion$-$nucleus 
scatterings, and indeed some indication of the increased in-medium 
pion$-$nucleon cross sections has been found in ref.\ \cite{lanl2}. 
Here we perform a rather exploratory study in order to set a limit on 
this exotic process. 
By simply increasing the pion$-$nucleon inelastic cross section by a factor of 
three without changing its branching ratios, we have found that the discrepancy 
at $T_{in}-T_{out}\geq 150$ MeV can be largely removed. As shown by the
dashed lines in Fig.\ 3, this modification only increases the cross section 
of produced pions without affecting the quasielastic parts of the spectra. 
In this respect, it is worthwhile to mention that several theoretical studies 
(e.g.\ \cite{brown1,brown2,kaplan,hatsuda,asakawa}) 
have shown that the properties of hadrons (e.g. $\Delta$ and $\rho$), 
such as their masses and decay widths, may be modified in the nuclear medium
as a result of the partial restoration of chiral symmetry. 
The appreciable effect on pion spectra due to increased pion$-$nucleon 
cross sections calls for more detailed study on medium effects.
More data of high energy pion$-$nucleus scatterings are thus desirable.

In summary, by using a hadronic transport model originally designed 
for heavy-ion
collisions we have analyzed the double differential cross sections 
in 500 MeV $C(\pi, \pi^{'})$ scatterings. The data
can be largely accounted for if one uses energy-dependent, anisotropic 
angular distributions that are fitted to pion$-$nucleon scattering data
in the decay of $\Delta(1232)$ and $N^*(1440)$ resonances.
We have also found that even with the careful inclusion of all presently known 
important pion$-$nucleon processes the semiclassical transport theory is not able 
to completely reproduce the pion$-$nucleus scattering data studied here. 
The disagreement of up to a factor of three in parts of the energy-loss 
spectra leaves open the possibility that exciting
new physics may play a role here.

Two of us (BAL and WB) would like to thank B. Jacak, M. Johnson and D.D. 
Strottman for inviting us to the Los Alamos workshop. We are grateful to 
J.D. Zumbro and W.R. Gibbs for many interesting discussions. 
We are also grateful to T.S.-H. Lee for providing us his code for calculating
consistently the angular distributions in $NN\leftrightarrow N\Delta$ 
processes. The work of BAL and CMK was supported in part by NSF Grant 
No. PHY-9212209 and PHY-9509266, and the work of WB was supported in part 
by NSF Grant No. PHY-9403666 and NSF Presidential Faculty Fellow grant,
PHY-9253505.

\newpage   
\section*{Figure Captions}
\begin{description}

\item{\bf Fig.\ 1 }\ \ \ 
	The angular distribution of pions from the decay of $\Delta^{++}$ 
resonance in its rest frame. 

\item{\bf Fig.\ 2 }\ \ \ 
	Pion absorption cross section as a function of beam energy
in $\pi^++^{12}C$ scatterings. The data are taken from refs. 
\cite{ashery,jones,ransome}. (see text). 

\item{\bf Fig.\ 3}\ \ \ Inclusive double differential cross
sections of pions as a function of the pion laboratory angle and the
pion energy loss, $T_{\rm in}-T_{\rm out}$, in 500 MeV $C(\pi^{+},\pi^{+'})$ 
reactions. The data (circles) are from ref. \cite{lanl1}.

\end{description}

\end{document}